\documentclass[twocolumn]{revtex4-2}
\usepackage{chemformula} 
\usepackage{verbatim}
\usepackage{graphicx}
\usepackage{grffile}
\usepackage{xcolor}
\usepackage{soul}
\usepackage{hyperref}
\graphicspath{{mainplots/}} 

\newcommand*\vc[1]{\mathbf{#1}}
\newcommand*\tx[1]{\mathrm{#1}}
\newcommand*\wn{cm$^{-1}$}
\newcommand*\ca{$\mathrm{C_{\alpha}}$}
\usepackage[symbol]{footmisc}

\newcommand{\1}[1]{\textcolor{blue}{#1}}

\newcommand{\0}[1]{}

\begin{document}
\title[]{High-Throughput Computation of Anharmonic Low-Frequency Protein Vibrations}

\author{Michael A. Sauer}
\author{Souvik Mondal}
\author{Madeline Cano}
\author{Matthias Heyden}
\email{mheyden1@asu.edu}
\affiliation{School of Molecular Sciences, Arizona State University, Tempe, AZ 85287, U.S.A.}

\date{\today}

\begin{abstract}
\normalsize
At room temperature, low frequency vibrations at far-infrared frequencies are thermally excited ($k_B T > h \nu$) and not restricted to harmonic fluctuations around a single potential energy minimum. 
For folded proteins, these intrinsically anharmonic vibrations can contain information on slow conformational transitions. 
Recently, we have developed FREquency-SElective ANharmonic (FRESEAN) mode analysis, a method based on time correlation functions that isolates low-frequency vibrational motions from molecular dynamics simulation trajectories without relying on harmonic approximations. 
We recently showed that low-frequency vibrations obtained from FRESEAN mode analysis are effective collective variables in enhanced sampling simulations of conformational ensembles.

However, FRESEAN mode analysis is based on velocity time correlations between all degrees of freedom, which creates computational challenges for large biomolecules. 
To facilitate future applications, we demonstrate here how coarse-graining of all-atom simulation trajectories can be combined with FRESEAN mode analysis to extract information on low-frequency vibrations at minimal computational cost.
\end{abstract}

\maketitle




\section{Introduction}
\label{s:intro}

Molecular dynamics (MD) simulations routinely capture atomic-level fluctuations and dynamic processes in proteins and other biomolecules on timescales ranging from femtoseconds to microseconds. 
However, protein conformational transitions typically occur on micro- to millisecond or even longer timescales, which remain challenging to sample in conventional MD simulations. 
To effectively capture protein dynamics in molecular dynamics simulations, enhanced sampling techniques have been developed to accelerate the sampling of protein conformational landscapes.

Currently, enhanced sampling methods can be broadly categorized into two main types: unbiased and biased approaches. 
Unbiased methods, such as Gaussian accelerated molecular dynamics \cite{GAMD} and replica exchange molecular dynamics \cite{REMD}, explore conformational space without predefined collective variables (CVs, also referred to as reaction coordinates) that describe the slow conformational transitions of a system. 
Biased methods, such as umbrella sampling \cite{Umbrella} or metadynamics \cite{METADYNAMICS}, rely on predefined CVs to guide sampling. 
The same applies to weighted ensemble methods \cite{zuckerman2017weighted}, which do not modify dynamic forces but accelerate sampling along CVs through duplication and merging of unbiased walkers.

However, the selection of suitable CVs is non-trivial and requires prior knowledge of the expected slow processes to be sampled.
If such information is not available, long unbiased simulation trajectories are often used to extract large amplitude motions or slow dynamics despite incomplete sampling.
For example, CVs associated with large amplitude motions can be extracted as eigenvectors of a co-variance matrix of fluctuations relative to an average structure\cite{David2014-PCA}.
Similarly, time-lagged independent component analysis (TICA) is based on displacement time correlations that identify CVs associated with slow dynamics \cite{Naritomi2013-TICA, Schultze2021}.
However, these methods rely on direct observations of rare events in long simulation trajectories and their reproducibility and throughput is limited.
Recent machine learning approaches provide potential alternatives through direct analysis of high-dimensional simulation trajectories to generate latent space representations that serve as CVs associated with underlying dynamic processes \cite{Tiwary2025, Majumder2024, Parrinello2018, Chen2022, chatterjee2025acceleration}. 

Here, we pursue a different strategy. 
Promising CVs that encode conformational transitions can also be obtained from low-frequency vibrations.
Low-frequency vibrational modes are often estimated using harmonic normal mode analysis, either for simplified structural representations of proteins such as elastic network models\cite{Bahar2005, Tirion1996, Bahar1997, Bahar1997V2}, atomistic protein structures\cite{NMA, Mahajan2015}, or structural ensembles\cite{romo2020persistent}.
Alternatively, fluctuations around an average structure observed in a dynamic simulation can be used to extract quasi-harmonic normal modes\cite{levy1984quasiA,levy1984quasiB}.
However, the implied harmonic or quasi-harmonic approximations are invalid for low-frequency vibrations, which are: 1) easily excited by thermal collisions; 2) not limited to fluctuations around a single potential energy minimum; 3) distorted by anharmonic effects, {\em i.e.}, the actual form of the potential.

To enable a correct characterization of low-frequency vibrations, we recently introduced FREquency-SElective ANharmonic (FRESEAN) mode analysis, which isolates low-frequency vibrations from MD simulation trajectories without the need for harmonic or quasi-harmonic approximations. 
FRESEAN mode analysis is based on a velocity time correlation approach that isolates degrees of freedom (DoF) contributing to the vibrational density of states (VDoS) at a given frequency. 
The sampling of velocity time correlation functions converges on a picosecond to nanosecond timescale and is highly reproducible in independent trajectories \cite{Sauer2024FastSample}. 

In previous work, we successfully demonstrated enhanced sampling of conformational dynamics using low-frequency anharmonic vibrations as CVs that were identified with FRESEAN mode analysis \cite{mondal2024exploring}. 
The test system in Ref.~\citenum{mondal2024exploring} has been alanine dipeptide with $N$=22 atoms.
FRESEAN mode analysis involves the calculation of time- and frequency-dependent velocity cross-correlations between all Cartesian DoF of the target molecule. 
For alanine dipeptide with $3N=66$ Cartesian DoF, this results in $66 \times 66 = 4,356$ matrix elements for each delay time in the correlation function and/or each sampled frequency. 
Obviously, for larger biomolecular systems, such as proteins and protein complexes with thousands or more atoms, the size of these matrices increases rapidly. 

Another concern is the time resolution required to sample the highest-frequency vibrations in the system.
Even when covalent bonds are constrained, 
which eliminates fast C-H, N-H, and O-H stretch vibrations,
atomic velocities need to be sampled at approximately 10~femtosecond intervals to avoid under-sampling.
Large trajectory files and high file I/O traffic caused by high sampling rates can be mitigated with data streaming approaches that enable the analysis of simulation data on the fly\cite{IMDClient}.
However, high sampling rates also increase the number of cross-correlation matrices that need to be stored in memory for a given maximum correlation time (typically a few picoseconds). 

In this work, we introduce strategies to minimize the computational overhead associated with FRESEAN mode analysis of molecular dynamics simulations of proteins.
We focus on the lowest-frequency vibrations in the system, which are often most relevant for applications associated with the identification of CVs for enhanced sampling simulations.
Using hen egg white lysozyme (HEWL) as a test system, we demonstrate that coarse-grained (CG) representations of all-atom simulation trajectories can retain almost all information on low-frequency vibrations.
The main requirement is that CG beads correspond to centers of mass (CoM) of groups of atoms. 
This approach filters out high-frequency vibrations between atoms contributing to a CG bead, while low-frequency vibrations between beads are preserved.
Such CG representations of all-atom trajectories greatly reduce the number of DoFs and the resulting size of correlation matrices used in FRESEAN mode analysis. 
At the same time, the absence of high-frequency vibrations allows for the sampling of velocities at larger time intervals, further reducing computational overhead and memory requirements. 
In general, our approach enables high-throughput analysis of anharmonic low-frequency vibrations in proteins and other biomolecules at minimal computational cost.

\section{Methods}
\label{s:methods}
\subsection{Simulation Protocol}
\label{s:simprotocol}
All simulations were performed using the GROMACS 2022.5 \cite{abraham2015} software package. 
In this study, HEWL (PDBID: 1HEL), a globular enzyme with previously studied dynamics (\cite{Levitt1985,Costa2015}), was selected to compare the results of FRESEAN mode analysis at different levels of coarse-graining. 
With 1960 atoms, HEWL is a protein with intrinsic collective dynamics that is still small enough to perform FRESEAN mode analysis on an all-atom level without the need of specialized hardware.

Notably, we performed all of the simulations in this study with all-atom resolution in explicit solvent. 
We then directly analyzed the all-atom trajectories and compared results to the analysis of CG representations of the same trajectories.

We used the AMBER99SB-ILDN force field\cite{amber99sb-ildn} for the protein in combination with the TIP3P water model\cite{jorgensen83}.
We placed the protein in the center of a periodic 100~\AA~$\times$~100~\AA~$\times$~100~\AA\ simulation box with 31,977 water molecules, 90 $\tx{Na}^{+}$ and 98 $\tx{Cl}^{-}$ ions (150~mM salt).
Our simulation protocol consisted of the following steps: 
First, we minimized the energy of the system using the steepest descent algorithm for 1000 steps. 
Then, we equilibrated the system in the isobaric--isothermal (NPT) ensemble at 300~K and 1~bar for 100~ps.
During equilibration, we used 1~fs integration timesteps, a velocity rescaling thermostat \cite{bussi2007} with a 1.0~ps time constant, and a stochastic cell rescaling barostat \cite{Bernetti2020} with a time constant of 2.0~ps.
This was followed by production simulations in the NPT ensemble with a 2~fs timestep.
Unless noted otherwise, results are reported for a single production simulation of 20~ns.
Only in the context of our statistical analysis, we further analyzed results from five replica simulations of 1~ns, 10~ns, 20~ns length as indicated.
In production simulations, we used a Nos\'{e}-Hoover thermostat \cite{nose1984,hoover1985} with a 1.0~ps time constant and a Parrinello--Rahman barostat \cite{parrinello1981} with a 2.0~ps time constant. 
All covalent bonds involving hydrogens were constrained using the LINCS\cite{hess1997} algorithm. 
Short-ranged electrostatic and Lennard-Jones interactions were treated with a 10~\AA~real--space cutoff with energy and pressure corrections for dispersion interactions. Long--ranged electrostatic interactions were treated with the Particle Mesh Ewald algorithm\cite{darden1993} using a 1.2~\AA~grid. 


\subsection{FRESEAN Mode Analysis}
\label{s:analysis}

The theory and computational steps associated with FRESEAN mode analysis are described in Ref.~\citenum{sauer2023} and are briefly summarized here.
The approach is agnostic to the representation of the system, {\em e.g.}, whether atomic velocities are analyzed or velocities of CG beads. 
As a first step in the analysis, we align the protein coordinates in every frame of the trajectory with an arbitrary reference structure using a combined translation and rotation operation on atomic coordinates that minimizes the root mean squared displacement (RMSD) relative to the reference.
The same rotation is applied to the protein velocities, which ensures that the following analysis of velocity correlations is performed in a consistent Cartesian frame of reference.
Notably, this rotation does not remove angular momenta.
To simplify the notation for mass-weighted velocity correlation functions, we introduce weighted velocities: $\tilde{\tx{v}_i}(t) = \sqrt{m_i} \cdot \tx{v}(t)$. 
Here, the index $i$ denotes one out of $3N$ Cartesian DoFs of the protein and $m_i$ describes the corresponding mass (atom or bead).
We then define a matrix $\vc{C}_{\tilde{\tx{v}}}(\tau)$ of velocity time correlation functions between all $3N$ DoFs that includes auto ($i = j$) and cross ($i\neq j$) correlations.
Individual matrix elements $C_{\tilde{\tx{v}},ij}$ are defined as:
\begin{equation}
C_{\tilde{\tx{v}},ij}(\tau)=\langle \tilde{\tx{v}}_{i}(t) \tilde{\tx{v}}_{j}(t+\tau)\rangle_{t}
\label{e:vccm}
\end{equation}
Here, $\left\langle ... \right\rangle_t$ denotes ensemble averaging over the simulation time and $\tau$ is the delay time between two simulation time frames.
Each time correlation function $C_{\tilde{\tx{v}},ij}(\tau)$ is calculated with a maximum delay time of $\tau_\tx{max}=$ 2~ps in steps ranging of 8 to 40~fs (see section~\ref{s:cgprotocol}).


After constructing 
$\vc{C}_{\tilde{\tx{v}}}(\tau)$, each element 
is Fourier transformed to obtain a frequency-dependent velocity correlation matrix $\vc{C}_{\tilde{\tx{v}}}(\omega)$:
\begin{equation}
    C_{\tilde{\tx{v}},ij}(\omega) = \int_{-\infty}^{+\infty}\tx{exp}(\tx{i}\,\omega \tau) C_{\tilde{\tx{v}},ij}(\tau) \,d\tau
    \label{eqn:vacf-ft}
\end{equation}

The sum of diagonal terms of $\vc{C}_{\tilde{\tx{v}}}(\omega)$ describes the sum of all Fourier transformed time auto-correlation functions and is directly proportional to the vibrational density of states (VDoS) \cite{chakraborty07,heyden10b,mathias11}.
\begin{equation}
\tx{VDoS}(\omega)=\frac{2}{k_\tx{B} T} \sum_i C_{\tilde{\tx{v}},ii}(\omega) 
\label{e:vdos1}
\end{equation}

The VDoS effectively describes the distribution of kinetic energy over vibrational frequencies (including zero frequency).
In equilibrium, normalization by the average kinetic energy per DoF ($k_\tx{B} T/2$) in Eq.~\ref{e:vdos1} ensures that the integral over all frequencies corresponds to the total number of protein DoFs.
\begin{equation}
    N_\tx{DoF} = \frac{1}{2\pi} \int_{0}^{+\infty}\tx{VDoS}(\omega)\,d\omega
    \label{e:int}
\end{equation}
For an all-atom representation of the protein with $N$ atoms and $N_H$ constrained covalent bonds involving hydrogen atoms, we obtain $N_\tx{DoF} = 3N-N_H$.

The frequency-dependent correlation matrix $\vc{C}_{\tilde{\tx{v}}}(\omega)$ can be diagonalized at any selected frequency $\omega_\tx{sel}$.
The sum of diagonal elements (trace) of $\vc{C}_{\tilde{\tx{v}}}(\omega)$ does not change upon diagonalization, which results in a simple relation between the sum of eigenvalues $\lambda_i=\lambda_i(\omega_\tx{sel})$ and the VDoS:
\begin{equation}
\tx{VDoS}(\omega_\tx{sel})=\frac{2}{k_\tx{B} T} \sum_i \lambda_i(\omega_\tx{sel})
\label{e:vdos2}
\end{equation}
This relation defines the meaning of the eigenvalues $\lambda_i$: they are directly proportional to VDoS contributions of mass-weighted velocity fluctuations along their corresponding eigenvector $\mathbf{q}_i$ at frequency $\omega_\tx{sel}$.
When applied to molecular systems, the vast majority of eigenvalues $\lambda_i$ are zero or close to zero for any given frequency.
The eigenvectors associated with the non-zero eigenvalues at $\omega_\tx{sel}$ provide a complete set of vibrational modes contributing to the VDoS at that frequency~\cite{sauer2023}.
Importantly, this approach does not rely on harmonic or quasi-harmonic approximations and can be applied to the lowest vibrational frequencies including zero frequency.
At zero frequency, the first six eigenvectors (largest eigenvalues) describe 
rigid-body translations and rotations (diffusion).
The remaining eigenvectors with non-zero eigenvalues describe low-frequency vibrations that contribute to the zero-frequency VDoS due to damping and/or anharmonicity~\cite{sauer2023}.
The frequency-dependent contributions of any given vibrational mode to the VDoS can be analyzed as described in the next section.


\subsection{VDoS Contributions of Individual Vibrational Modes}
\label{s:1dvdos}
To quantify contributions of a single vibrational mode $\vc{q}_i$, {\em e.g.}, one of the eigenvectors of $\vc{C}_{\tilde{\tx{v}}}(\omega)$ at a given frequency $\omega_\tx{sel}$, to the overall VDoS, we first project weighted velocities ($\tilde{\tx{v}}(t)$) of all 3N degrees of freedom onto $\vc{q}_i$.
\begin{equation}
\dot{q}_i(t) = \sum_{j} \tilde{\tx{v}}_{j}(t) \, q_{ij}
\label{e:qdot}
\end{equation}
The projected velocities $\dot{q}_i(t)$ describe the system's dynamics along $\vc{q}_i$. 
The Fourier transform of the corresponding time auto-correlation function describes contributions of fluctuations along $\vc{q}_i$ to the VDoS at all sampled frequencies, {\em i.e.}, the VDoS for the single (collective) DoF corresponding to $\vc{q}_i$.

\begin{eqnarray}
C_{q_i}(\tau) &=&\langle \dot{\tx{q}}_i(t) \dot{\tx{q}}_i(t+\tau)\rangle_t \\
\tx{VDoS}_{\vc{q}_i}(\omega) &=& \frac{2}{k_\tx{B} T} \int_{-\infty}^{+\infty} \tx{exp}(\tx{i}\,\omega \tau) C_{q_i}(\tau)\,d\tau
\label{e:qvdos}
\end{eqnarray}

We note that this information is also encoded in $\vc{C}_{\tilde{\tx{v}}}(\omega)$. 
As a consequence, we can alternatively compute $\tx{VDoS}_{\vc{q}_i}(\omega)$ directly as:

\begin{equation}
\tx{VDoS}_{\vc{q}_i}(\omega) = \frac{2}{k_\tx{B} T} \, \vc{q}_i^T \vc{C}_{\tilde{\tx{v}}}(\omega) \vc{q}_i
\end{equation}

Notably, if $\vc{q}_i$ is the $i$'th eigenvector of $\vc{C}_{\tilde{\tx{v}}}(\omega)$ at frequency $\omega_\tx{sel}$ then $\tx{VDoS}_{\vc{q}_i}(\omega_\tx{sel}) = \frac{2}{k_\tx{B} T} \lambda_i(\omega_\tx{sel})$.


\subsection{CG Representations of All-Atom Trajectories}
\label{s:cgprotocol}

\begin{figure}[ht!]
\begin{center}
\includegraphics[clip=true,width=1\linewidth]{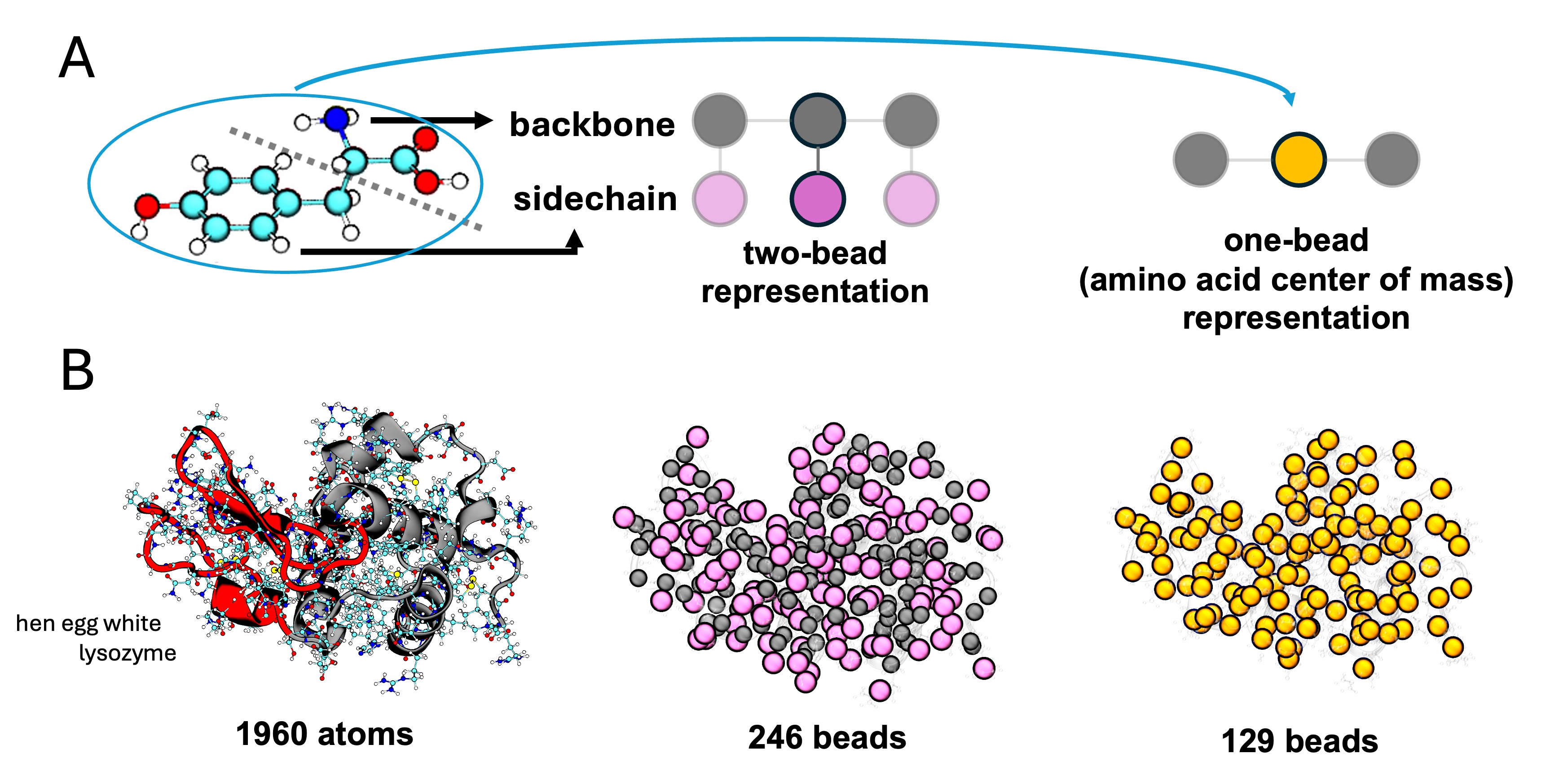}
\caption{Coarse-graining schemes A: Mapping of the center-of-mass of backbone and side-chain atoms of tyrosine (left) into separate CG beads (middle) or a single bead (right).
B: Visualization of all-atom and CG representations of HEWL with 1960 atoms, 246 beads, 129 beads, respectively.
}
\label{f:cg}
\end{center}
\end{figure}

We used three distinct CG representations of all-atom HEWL trajectories to evaluate how much information on low-frequency vibrations is preserved upon decreasing the number of DoFs. 
In the first representation (two-bead, see Figure~\ref{f:cg}), the CoM positions and velocities of backbone and side chain atoms of each amino acid residue are mapped to separate beads (two beads per residue). 
Glycine is an exception to this rule for which the CoM of all residue atoms is mapped to a single bead.
Atoms for the N- and C-termini are included in the mapping to the backbone beads of the first and last residue, respectively.
For HEWL, this representation reduces the number of particles from 1960 atoms to 246 beads.

In the second CG representation (one-bead), the CoM of all atoms of a residue is mapped to a single bead (Figure~\ref{f:cg}), and the third representation uses a single bead corresponding to the \ca atom of each amino acid residue.
Both representations reduce the number of particles further to 129.


In addition to reducing the number of particles and DoF, we further tested distinct time resolutions for the calculation of the correlation functions.
A time resolution of $\Delta t=$8~fs allows sampling frequencies of up to 2085~\wn, which, in a simulation with constraints applied to covalent bonds involving hydrogens, includes the fastest vibrations expected in the system.
With increasing degrees of coarse-graining, especially when using centers-of-mass to define the mapping to CG beads, we further eliminate high-frequency vibrations. 
This allows us to further decrease the time resolution without under-sampling fluctuations, which would otherwise cause aliasing during Fourier analysis. 
Specifically, we tested reduced time resolutions of 20~fs and 40~fs. 

\subsection{Computational Costs and Memory Requirements}
\label{s:comp}
Both the computational overhead and memory requirements decrease substantially upon reducing the number of particles $N$ and the time resolution.
The construction of $\vc{C}_{\tilde{\tx{v}}}(\omega)$ requires the calculation of $\mathcal{O}(3N)^2$ time correlation functions and their Fourier transforms. 
In its current implementation~\cite{FRESEAN,FRESEANtutorial}, the calculation of the time correlation functions utilizes fast Fourier transforms of full-length time-dependent velocity trajectories ($t_\tx{sim}$ = total simulation time) and a variation of the convolution theorem:
\begin{eqnarray}
    \tx{v}_i(\omega) &=& \int_0^{t_\tx{sim}} \tx{exp}(\tx{i}\,\omega t) \tx{v}_i(t) \, dt \\
    C_{\tilde{\tx{v}},ij}(\omega) &=& \tx{v}_i(\omega) \overline{\tx{v}_j(\omega)}
    \label{e:conv}
\end{eqnarray}
Here, $\overline{\tx{v}_j(\omega)}$ indicates the complex conjugate. 
The frequency resolution is then adjusted {\em a posteriori} via a convolution with a square window function in the time domain that is non-zero between $t=0$ and $\tau_\tx{max}$.

While this approach requires storing trajectories of all particle velocities in memory, the dominant memory requirement for large numbers of particles $N$ typically results from storing the $(3N)^2$ correlation functions.
Further, simulation trajectories can be easily analyzed in shorther fragments and resulting correlation matrices $\vc{C}_{\tilde{\tx{v}}}(\omega)$ can be averaged afterwards.

The key advantage of the approach in Eq.~\ref{e:conv}, for large or small numbers of particles alike, is that the computational cost of calculating each correlation function is reduced from $\mathcal{O}(n_t \times n_\tau)$ to $\mathcal{O}(n_t\,\tx{log}\,n_t)$ (with time resolution $\Delta t$; $n_\tau = \tau_\tx{max} / \Delta t$; and $n_t = t_\tx{sim} / \Delta t$).

For large number of particles or if many frequencies are being analyzed, the computational cost of calculating the eigenvalues and eigenvectors of $\vc{C}_{\tilde{\tx{v}}}(\omega)$ is also worth considering, which scales with $\mathcal{O}(3N)^3$ and thus decreases dramatically upon coarse-graining.


\subsection{Comparing Vibrational Modes in Distinct Representations}
\label{s:similarity}

FRESEAN mode analysis can be applied to time-dependent trajectories of particle velocities irrespective of whether all-atom or CG representations are used. 
However, the number of DoF dictates the dimensionality of eigenvectors computed from $\vc{C}_{\tilde{\tx{v}}}(\omega)$ ({\em i.e.}, their number of components).
If two sets of eigenvectors contain the same number of components ({\em e.g.}, eigenvectors obtained from the same trajectory but for different frequencies, or eigenvectors at the same frequency obtained from distinct trajectories), we can compare them using their (unsigned) dot product.
Since the eigenvectors are normalized, the dot product can be interpreted as the cosine of the angle between both vectors (or its equivalent in a high-dimensional space).

\begin{equation}
\tx{cos}(\theta_{ij}) = \left| \vc{q}_i \cdot \vc{q}_j \right|
\label{e:similarity}
\end{equation}

When $\tx{cos}(\theta_{ij})=1$, the eigenvectors are identical, while $\tx{cos}(\theta_{ij})=0$ indicates that two vectors are orthogonal and share no information ({\em e.g.}, distinct eigenvectors computed from the same matrix).

However, when comparing results from all-atom and CG representations, we need to compare eigenvectors with distinct dimensionality.
For this comparison, for which we only consider the two-bead and one-bead representations of residues, we translate all-atom eigenvectors into a CG representation that converts atomic displacements into the corresponding CoM-displacements.

Thus, if $\vc{q}_{ik}^\tx{aa}$ is the $k$'th atomic displacement vector of the $i$'th all-atom eigenvector $\vc{q}_{i}^\tx{aa}$, then the displacement vector of the $j$'th bead in the corresponding CG representation of that eigenvector is given by:
\begin{equation}
\vc{q}_{ij}^\tx{cg} = \frac{\sum_{k \in \tx{bead}_j} m_k \, \vc{q}_{ik}^\tx{aa}}{\sum_{k \in \tx{bead}_j} m_k}
\label{e:dim}
\end{equation}


\section{Results and Discussion}
\label{s:results}

\subsection{Frequency-Dependent Information in CG Representations}

\begin{figure*}[ht!]
\includegraphics[clip=true,width=1\linewidth]{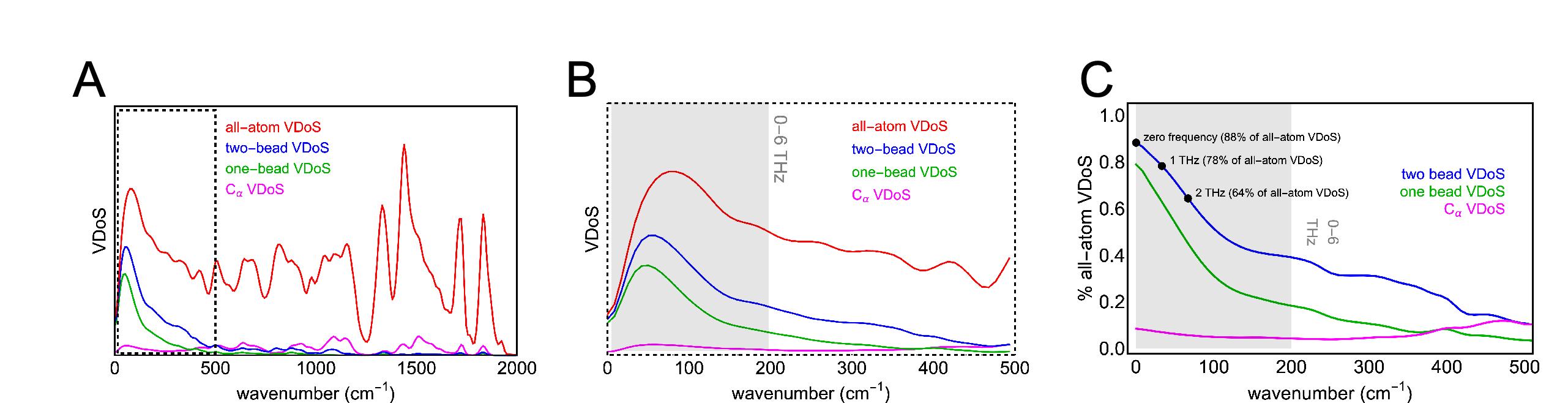}
\caption{
(A) Vibrational density of states (VDoS) of HEWL in the all-atom (red), two-bead (blue), one-bead (green), and \ca\ (magenta) representations for all frequencies sampled in the simulation.
The boxed region highlights the frequency range below 500~\wn\ shown as a close-up in (B). 
A gray background in (B,C) highlights the 0–6 THz range, where the thermal energy exceeds harmonic oscillator excitation energies ($k_\tx{B} T \geq h \nu$).
(C) Fraction of the all-atom VDoS captured by the CG representations as a function of frequency.
}
\label{f:vdos}
\end{figure*}

In Figure~\ref{f:vdos}, we compare the VDoS of HEWL obtained from all-atom, two-bead, one-bead, and \ca-atom representations of the all-atom simulation trajectories.
All trajectories were analyzed with a time resolution of 8~fs.
As shown in Eq.~\ref{e:int}, the integral of the VDoS over all frequencies corresponds to the total number of DoF, which is inherently different for each representation: the all-atom system has 4921 DoF (3 DoF for each of the 1960 atoms minus 959 constrained bonds), while the CG representations retain only 738 DoF (two-bead) and 387 DoF (one-bead or \ca), respectively.

The all-atom VDoS exhibits a peak in the far-IR region (50-80~\wn), consistent with previous inelastic neutron scattering and molecular dynamics results \cite{Lerbret2012}. 
This peak occurs at a frequency similar to intermolecular vibrations in the hydrogen bond network of the surrounding hydration shell \cite{heyden2013spatial}. 
In the fingerprint and mid-IR regions, intramolecular bond vibrations between heavy (non-hydrogen) atoms of the protein are observed. 
However, resonances beyond 2000~\wn\ are absent due to constrained bonds involving hydrogen atoms. 

For the two-bead and one-bead representations, vibrations are almost entirely restricted to frequencies below 500~\wn. 
The center-of-mass representations largely eliminate vibrations between individual atoms.
In contrast, vibrations between separate beads, which correspond to collective vibrations in the protein, are retained and observed at low frequencies.

Notably, we obtain a very different VDoS for the \ca\ atoms of the protein. 
The \ca\ atoms participate in intramolecular bond vibrations, which contribute to the spectrum throughout the entire frequency range.  
Therefore, \ca\ atom-based CG representations are not well suited to isolate collective low-frequency vibrations in the protein.

In Figure~\ref{f:vdos}B, we zoom in on the VDoS at frequencies below 500~\wn\ and highlight the region below 6~THz ($\approx$~200~\wn; 1~THz~$\approx$~33~\wn), where the thermal energy (at 300~K) exceeds harmonic oscillator excitation energies and prominent anharmonic effects can be expected.
The comparison of the VDoS obtained for the different CG representations clearly illustrates the frequency-dependent information loss in each of them.
Most notably, the two-bead and one-bead representations trace the all-atom VDoS for the lowest frequencies up to several tens of \wn. 
Due to the higher spatial resolution, the two-bead representation reproduces the all-atom VDoS slightly better than the one-bead representation. 
However, at zero frequency both the two-bead and one-bead representations reproduce the all-atom VDoS accurately (in contrast to the \ca\ representation).

This is further highlighted in Figure~\ref{f:vdos}C, where we plot the ratio of the VDoS obtained from the CG and all-atom representations. 
At zero frequency, the two-bead representation captures 88\% of the all-atom VDoS, which drops to 78\% and 65\% for 1 and 2~THz, respectively.
The one-bead representation captures around 80\% of the zero-frequency VDoS intensity, which decreases with increasing frequency analogously to the two-bead representation.
In contrast, the \ca\ representation captures only $\approx$~10\% of the all-atom VDoS at zero-frequency.

The results show that both the two-bead and one-bead representations reproduce a substantial fraction of collective protein vibrations for the lowest frequencies.
Nevertheless, the fraction of retained information drops rapidly with increasing frequency.
If one is interested in protein vibrations at frequencies beyond 1-2~THz, the all-atom representation is essential, {\em e.g.}, to capture low-frequency vibrations associated with dihedral angle rotations, etc. 
However, our previous work on the extraction of CV candidates for enhanced sampling simulations focused entirely on vibrations contributing to the VDoS at zero frequency~\cite{mondal2024exploring,Sauer2024FastSample}, for which the two-bead and one-bead representations appear promising.
In the following, we will thus quantify how much coarse-graining affects the characterization of these vibrations.

Reproducing a large fraction of the all-atom VDoS at zero frequency does not yet prove that the corresponding CG vibrations faithfully reproduce the corresponding collective all-atom vibrations. 
A significant fraction of the zero-frequency VDoS is associated with rigid-body translations/rotations and of limited interest in the context of vibrations.
In addition, reproducing the magnitude of fluctuations does not necessarily imply that the underlying motions are identical.
Thus, in the following section, we directly compare the eigenvectors obtained for all-atom and CG representations of our simulations using FRESEAN mode analysis. 
These eigenvectors describe collective motion in terms of displacement vectors relative to the reference structure used during the translational/rotational alignment of the trajectories.

 
\subsection{Eigenvector Comparison}

\begin{figure*}[ht!]
\begin{center}
\includegraphics[clip=true, width=1\linewidth]{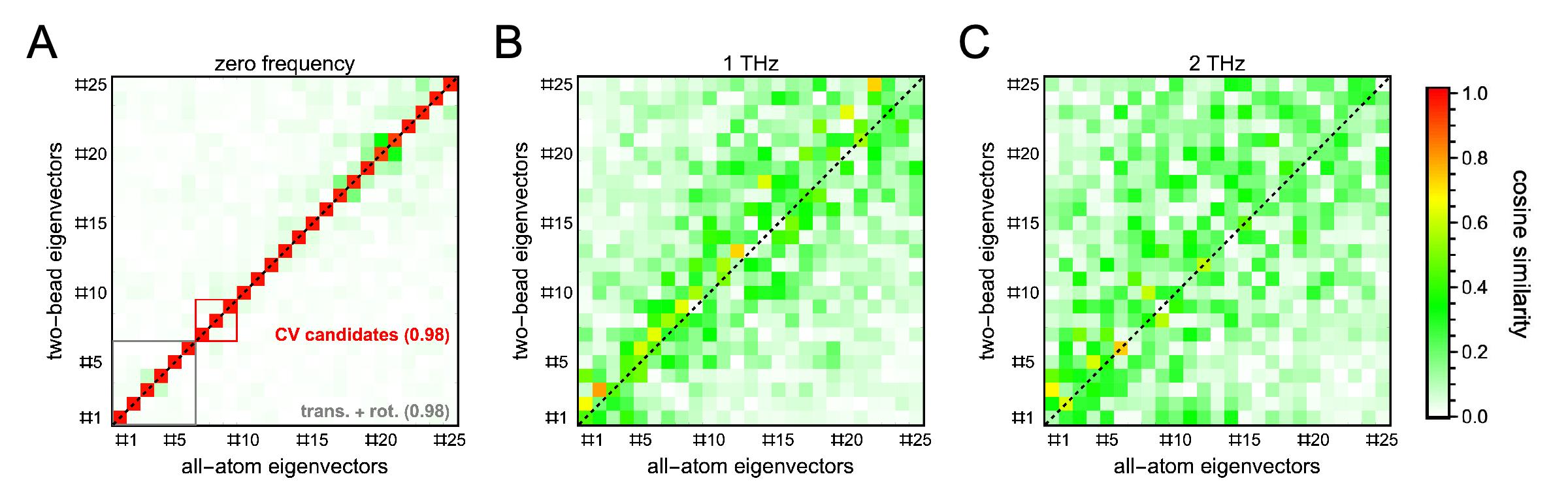}
\caption{Cosine similarity matrices between the first 25 eigenvectors of the velocity correlation matrix $\vc{C}_\tx{\tilde{v}}(\omega)$ at zero frequency (A), 1~THz (B), and 2~THz (C). 
Each matrix compares eigenvectors sampled from 20 ns simulations. 
A cosine similarity of 1 indicates identical eigenvectors, while a value of 0 indicates orthogonal eigenvectors. 
At zero frequency, we highlight eigenvectors describing rigid-body translations and rotations (eigenvectors 1-6, gray box) and vibrations with large contributions to the zero frequency VDoS (eigenvectors 7, 8, and 9, red box), which we identified as potential CV candidates for enhanced conformational sampling in previous work~\cite{Sauer2024FastSample}. Average cosine similarities along the diagonal within each box are indicate as numerical values.}
\label{f:vec}
\end{center}
\end{figure*}

In Figure~\ref{f:vec}, we show comparisons between eigenvectors of the velocity correlation matrices $\vc{C}_\tx{\tilde{v}}(\omega)$ obtained from all-atom and two-bead CG representations at zero frequency, 1~THz, and 2~THz. 
An analogous comparison for the one-bead representation is shown in Figure~S1 in the Supplementary Information (SI).
Comparisons are visualized as matrices of cosine similarities between eigenvectors obtained from trajectories in both representations. 
The eigenvectors are sorted by descending eigenvalues, {\em i.e.}, the eigenvector with index 1 corresponds to the largest eigenvalue of the matrix $\vc{C}_{\tilde{\tx{v}}}(\omega_\tx{sel})$.
To compare vectors of distinct dimensionality, we used the approach described in Eq.~\ref{e:dim}.

The comparison in Figure~\ref{f:vec}A shows clearly that zero-frequency eigenvectors are essentially equivalent in the all-atom and two-bead representations.
The cosine similarities of eigenvectors along the diagonal (eigenvectors with the same index) are on average 0.98.
This observation is independent of whether we compare eigenvectors 1-6, which describe rigid-body translations and rotations (see Figure~S2 of the SI for visualization), or higher-index modes that correspond to vibrations. 
We specifically highlight eigenvectors 7, 8, and 9 at zero frequency, which describe vibrational modes contributing to the zero-frequency VDoS.
In previous work, we identified these orthogonal vibrational modes as CV candidates for enhanced sampling simulations~\cite{mondal2024exploring,Sauer2024FastSample}.

In FRESEAN mode analysis, increasing mode indices (decreasing eigenvalues; see Figure~S3 of the SI) correspond to decreasing contributions to the VDoS at the selected frequency.
When eigenvalues become close or equal to zero, their eigenvectors are degenerate, {\em i.e.}, any linear combination of these eigenvectors describes a collective DoF with no contributions to the VDoS at the selected frequency.
Thus, with increasing index, we expect eigenvectors to become increasingly randomized.
Nevertheless, the high similarity between all-atom and CG eigenvectors in the two-bead representation extends beyond index 25.
Notably, our results for the one-bead representation in Figure~S2 of the SI are remarkably similar. 
This shows that the zero-frequency VDoS is dominated by collective motions in which all atoms within a given residue move together.

However, eigenvector similarities decrease substantially for non-zero frequencies as shown in Figure~\ref{f:vec}B and C. 
While frequencies of 1 and 2~THz are still deep in the far-infrared and typically also associated with collective vibrations, the lack of atomistic detail in the two-bead (and one-bead) representation limits cosine similarities along the diagonal of the comparison matrices to values around 0.5. 
We find different ordering for several eigenvectors, {\em i.e.}, the CG eigenvector with the highest similarity to a given all-atom eigenvector has a different index and is thus not found along the diagonal in the comparison matrix.
Several entries of similar magnitude in a column of the comparison matrix further indicate that the corresponding all-atom eigenvector is best described by a linear combination of multiple CG eigenvectors.

\subsection{VDoS of Eigenvector Projections}

\begin{figure}[ht!]
\includegraphics[clip=true, width=1.0\linewidth]{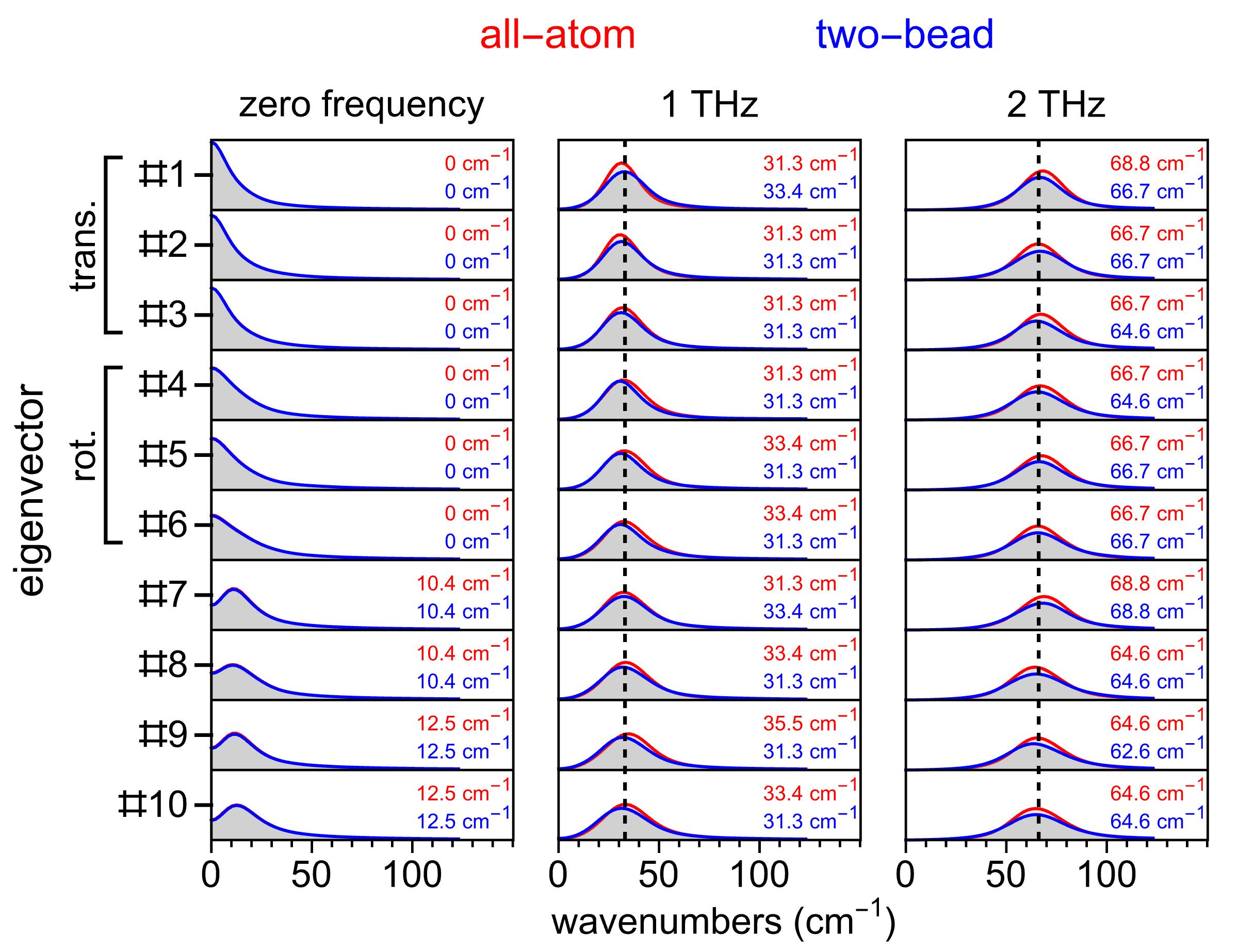}
\caption{1D-VDoS from fluctuations along the first 10 eigenvectors of $\vc{C}_\tx{\tilde{v}}(\omega)$ at zero frequency, 1~THz and 2 THz obtained for all-atom (red) and two-bead (blue) representations of protein trajectories. 
Colored labels indicate peak frequencies in each spectrum and vertical dashed lines indicate the frequency for which $\vc{C}_\tx{\tilde{v}}(\omega)$ was evaluated.}
\label{f:1dvdos}
\end{figure}

In Figure \ref{f:1dvdos}, we analyze the spectrum of velocity fluctuations projected on the first 10 eigenvectors of $\vc{C}_\tx{\tilde{v}}(\omega)$ at zero frequency, 1~THz and 2~THz for all-atom and two-bead CG (two-bead) representations of our simulation trajectories (see Figure~S4 of the SI for an analogous analysis with the one-bead representation).
The displayed spectra represent VDoS contributions from fluctuations along individual collective degrees of freedom, as defined by their respective eigenvectors (1D-VDoS, see Eq.~\ref{e:qvdos}).

As mentioned earlier, the first six eigenvectors at zero frequency (left panel) describe rigid-body translations and rotations. 
The 1D-VDoS for these diffusive motions feature a maximum at zero frequency that is directly proportional to the corresponding translational and rotational diffusion coefficients~\cite{zwanzig1965time}.
The 1D-VDoS for zero-frequency eigenvectors 7-10 describe collective low-frequency vibrations of the protein with peaks close to 10~\wn.
The spectra also show why these vibrational modes were isolated by our analysis: the broad lineshape of the peaks result in significant VDoS contributions at zero frequency (and thus non-zero eigenvalues of $\vc{C}_\tx{\tilde{v}}(\omega)$ at zero frequency).
The broad lineshape can be attributed to the anharmonicity of the anharmonic potential and/or damping, {\em e.g.}, due to energy dissipation into the solvent.

The most critical observation in the context of our analysis is that fluctuations along zero-frequency eigenvectors result in essentially identical 1D-VDoS for all-atom and CG representations (the same is true for the one-bead representation as shown in Figure~S4 of the SI).
This confirms that the two-bead and one-bead representations of all-atom protein simulations preserve information on the lowest-frequency vibrations.

As already seen in panels B and C of Figure~\ref{f:vec}, this is not the case for eigenvectors of $\vc{C}_\tx{\tilde{v}}(\omega)$ at frequencies $\ge$ 1~THz.
In the central and right panels of Figure~\ref{f:1dvdos}, we plotted the 1D-VDoS for fluctuations along the first 10 eigenvectors obtained at 1 and 2~THz.
Since we already observed in Figure~\ref{f:vec} that the eigenvectors in the all-atom and CG representations are not equivalent, it is not surprising that spectra of fluctuations along the eigenvectors are not identical either.

The eigenvectors are selected by our analysis for their VDoS contributions at 1 and 2~THz.
Consequently, we observe peak intensities close to both frequencies irrespective of the trajectory representation.
However, we observe minor differences in peak frequencies and increased peak intensities for fluctuations along eigenvectors obtained from the all-atom representation.
The integral of each VDoS describes the kinetic energy of a single DoF (see Eq.~\ref{e:int}). 
The increased peak intensities in the all-atom 1D-VDoS thus correspond to a narrower spectrum that is broadened by the CG representation.

The CG eigenvectors at 1 and 2~THz likely still contain valid general information on far-infrared protein vibrations. 
However, in contrast to zero-frequency CG eigenvectors, which reproduce all-atom motions essentially exactly, the information loss for 1 and 2~THz vibrations due to coarse-graining is non-negligible.

As mentioned earlier, we identified zero-frequency eigenvectors associated with lowest-frequency vibrations as CV candidates for enhanced sampling in previous work~\cite{mondal2024exploring,Sauer2024FastSample}.
For such applications, both the two-bead and one-bread CG representations are ideal as they preserve all necessary information at a fraction of the computational cost and memory requirements.

\subsection{Reducing Time Resolution} 

\begin{figure}[ht!]
\begin{center}
\includegraphics[clip=true,width=1\linewidth]{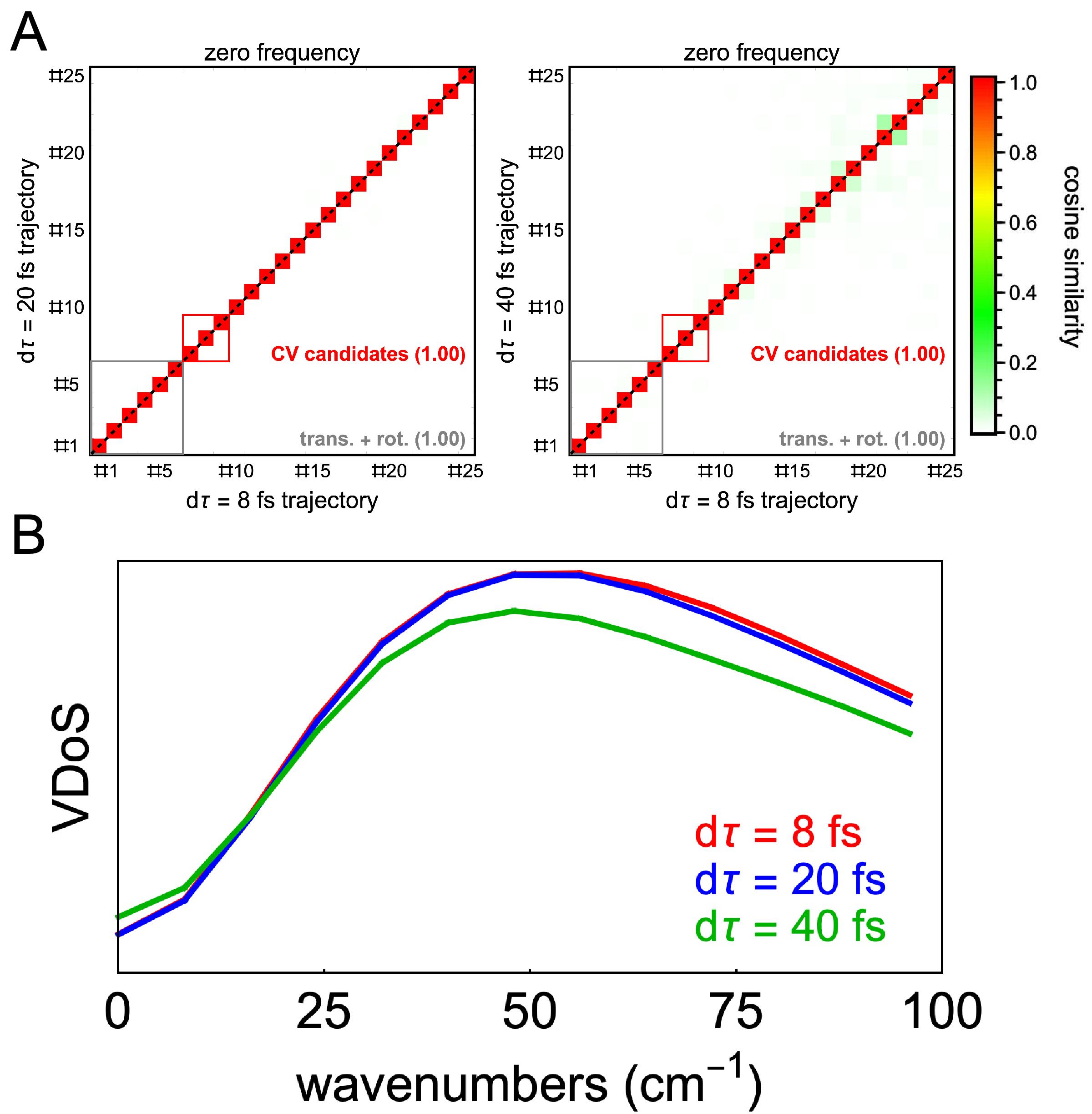}
\caption{(A) Cosine similarity matrices between the first 25 eigenvectors of the velocity correlation matrix $\vc{C}_\tx{\tilde{v}}(\omega)$ obtained with the two-bead representation at time resolutions of 8~fs and 20~fs (left) and 8~fs and 40~fs (right).
(B) VDoS obtained for the two-bead representation of trajectories with time resolutions of $d\tau$ = 8 (red), 20 (blue), and 40 fs (green).}
\label{f:dt}
\end{center}
\end{figure}

In Figure~\ref{f:dt}A, we compare zero frequency eigenvectors obtained from the two-bead representation of our trajectories with time resolutions of 8~fs, 20~fs, and 40~fs, while we compare the corresponding total VDoS in Figure~\ref{f:dt}B. 
The highest frequencies sampled at each of these time resolutions correspond to 2085~\wn, 834~\wn, and 417~\wn, respectively (Nyquist's theorem). 
A comparison to the VDoS of the two-bead representation in Figure~\ref{f:vdos} shows that a 20~fs time resolution only cuts off minor signals beyond 800~\wn. 
It is thus not surprising that a reduced time resolution of 20~fs has no effect on the eigenvectors of $\vc{C}_\tx{\tilde{v}}(0)$ and the total VDoS.

However, a time resolution of 40~fs under-samples vibrations around 400~\wn\ that contribute to a high-frequency shoulder of the main VDoS peak.
Consequently, we observe small but detectable discrepancies to the all-atom simulation for eigenvectors with index 15 and beyond.
Under-sampling high frequency vibrations results in false signals at low frequencies (aliasing), which we observe in the form of increased zero-frequency VDoS in Figure~\ref{f:dt}B. 
At the same time, VDoS intensities for frequencies $>$25~\wn\ are diminished due to a shift to higher frequencies.

Here, the one-bead representation should be more forgiving.
As shown in Figure~\ref{f:vdos}, the VDoS obtained from the one-bead representation has only minor contributions beyond 400~\wn. 
As shown in Figure~S5 of the SI, the zero-frequency VDoS for a time resolution of 40~fs is now essentially identical to the results for time resolutions of 8~fs and 20~fs. 
However, the eigenvectors and VDoS beyond 25\~wn\ show similar deviations between time resolutions of 8~fs and 40~fs time resolutions as observed for the two-bead model.
Overall, the one-bead representation is only slightly less sensitive to the time resolution of the trajectory compared to the two-bead model.

In both cases, a lower time resolution decreases the number of correlation time and frequency points that must be stored for each matrix element of $\vc{C}_\tx{\tilde{v}}(\tau)$ and $\vc{C}_\tx{\tilde{v}}(\omega)$. 
This further reduces computational cost and memory requirements for FRESEAN mode analysis in addition to coarse-graining.

\subsection{Visualization of Low-Frequency Vibrations}

\begin{figure}[ht!]
\begin{center}
\includegraphics[width=1\linewidth]{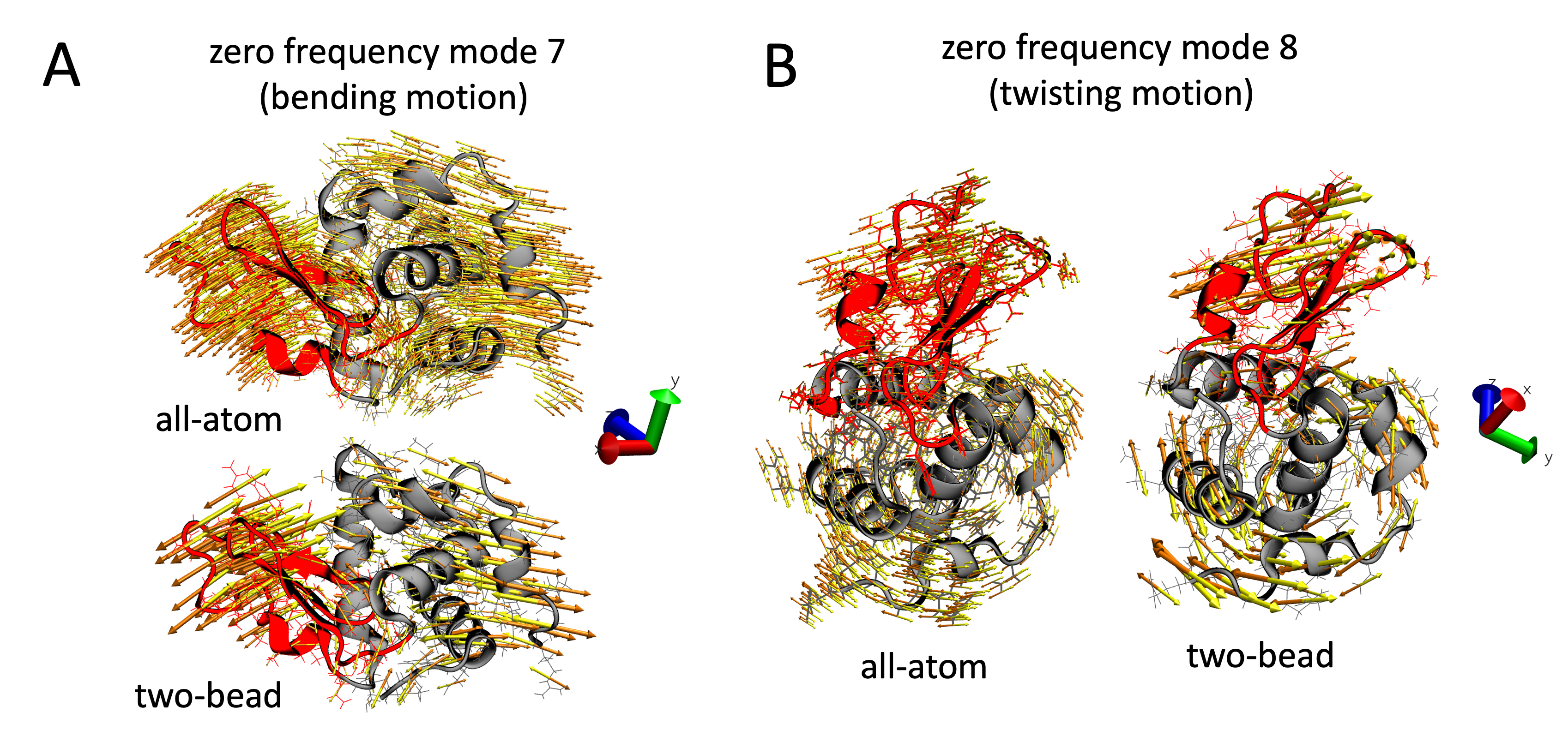}
\caption{Visualization of eigenvectors 7 and 8 of $\vc{C}_\tx{\tilde{v}}(\omega)$ at zero frequency obtained from all-atom and two-bead representations of our simulation trajectories. 
Displacement vectors are shown as bi-directional arrows that are superimposed on a reference structure. Orange (yellow) arrows describe the positive (negative) phase of an oscillation. 
The reference structure is shown in a cartoon representation to highlight secondary structure elements.
The $\alpha$-domain is shown in gray and the $\beta$-domain is highlighted in red.}
\label{f:modes}
\end{center}
\end{figure}

In Figure~\ref{f:modes}, we visualized eigenvectors 7 and 8 of $\vc{C}_\tx{\tilde{v}}(\omega)$ at zero frequency in their all-atom and two-bead representations.
We visualized the atomic and CG-bead displacements corresponding to the collective vibration as bi-directional arrows (positive and negative phase of the vibration) that are superimposed on the reference structure used for translational/rotational alignment of the trajectories.
As shown in Figure~\ref{f:1dvdos}, fluctuations along both eigenvectors contribute a low-frequency peak to the VDoS with a peak frequency of 10.4~\wn. 
Visual inspection confirms the high level of similarity of collective motions described on the all-atom and CG level.
The two vibrations correspond to distinct large-scale deformations that can be compared to similar motions observed using harmonic normal modes (\cite{Costa2015}, \cite{mackerell98}). 
However, the differences between harmonic normal modes and the vibrational modes identified via FRESEAN mode analysis matter: as shown in previous work on the Trp-cage protein~\cite{sauer2023}, harmonic and quasi-harmonic normal modes fail to identify collective DoF that isolate low-frequency vibrations, while the 1D-VDoS in Figure~\ref{f:1dvdos} each exhibit only a single low-frequency peak.

Eigenvector 7 indicates a concerted opening/closing motion of the $\beta$-domain (residues 40-88; shown in red) against the $\alpha$-domain (residues 1-39 and 89-129; shown in gray)~\cite{krebs2000formation}, which can alternatively be described by the pincer angle between both domains\cite{Costa2015}. 
Eigenvector 8 indicates an orthogonal motion, where both $\beta$-domain and $\alpha$-domains twist in opposite directions around a shared axis (parallel to direction of view in Figure~\ref{f:modes}B).
We successfully employed both eigenvectors in previous work as CVs for enhanced sampling simulations of HEWL conformations~\cite{Sauer2024FastSample}.



\subsection{Statistical Reproducibility}

In addition to coarse-graining of all-atom trajectories and selecting the time resolution, another factor that determines the computational cost of FRESEAN mode analysis is the length of the simulation time required to obtain converged results. 

To determine the optimal length of simulation trajectories, we tested simulation times of 1~ns, 10~ns and 20~ns and repeated the simulation protocol described in section~\ref{s:simprotocol} five times with starting velocities sampled independently from a Maxwell-Boltzmann distribution at 300~K.
We aim to determine the simulation time needed to obtain reproducible eigenvectors from the correlation matrix $\vc{C}_\tx{\tilde{v}}(\omega)$ at zero frequency in simulations starting from a given starting structure (here: crystal structure with PDBID: 1HEL).
For this analysis, we used the two-bead representation of all-atom trajectories with a time resolution of 20~fs (see Figure~S6 in the SI for analogous analysis with the one-bead model).

\begin{figure*}[ht!]
\begin{center}
\includegraphics[clip=true,width=0.8\linewidth]{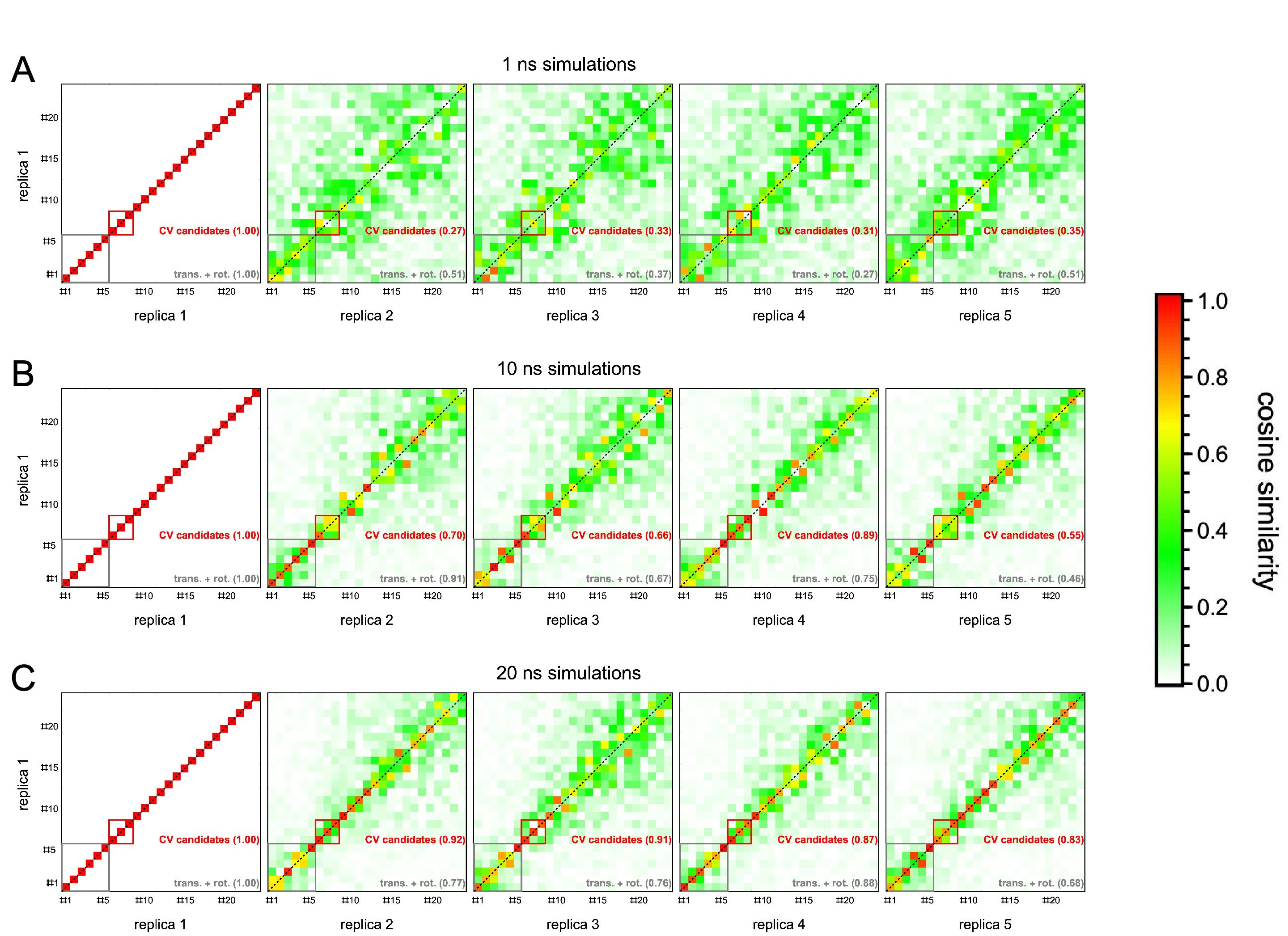}
\caption{Reproducibility of eigenvectors of $\vc{C}_\tx{\tilde{v}}(\omega)$ at zero frequency obtained from a two-bead representation of five replica all-atom simulations trajectories of 1~ns (A), 10~ns (B), and 20~ns (C) length.
All simulations are compared to replica 1 for clarity and shown are matrices of cosine similiarities between pairs of the first 25 eigenvectors.
Eigenvectors describing rigid-body translations and rotations (1-6) and the lowest frequency vibrations (7-9) are highlighted as in Figures~\ref{f:vec} and \ref{f:dt}.}
\label{f:reprod}
\end{center}
\end{figure*}

To quantify reproducibility, we again use cosine similarity matrices to compare eigenvectors generated in five replica simulations for each trajectory length as shown in Figure \ref{f:reprod}.
Instead of an all-to-all comparison, we compare eigenvectors obtained from replicas 1-5 with replica 1 to simplify the visualization.
We include the trivial comparison of replica 1 with itself for clarity.

In Figure~\ref{f:reprod}A, we show the results for 1~ns simulations, which demonstrate substantial statistical variations. 
While general correlations exist between eigenvectors obtained from distinct simulations along the diagonal, the cosine similarity mostly remains below 0.6.
We occasionally observe higher correlations off-diagonal, in particular for the first 6 eigenvectors that describe rigid-body translation and rotation. 
The latter indicates similar eigenvectors but a distinct ordering of the corresponding eigenvalues.
However, we focus our attention primarily on eigenvectors 7-9 that describe the lowest-frequency vibrations of the protein. 
Here, the average correlation between eigenvector pairs with identical indices ranges only from 0.27 to 0.35, indicating substantial variations between replicas.

For 10~ns simulation in Figure~\ref{f:reprod}B, the reproducibility improves significantly.
We still observe distinct ordering of eigenvectors describing translations and rotations. 
However, especially eigenvector 7, which describes the lowest frequency vibration of the protein, is reproduced with high confidence.
Compared to replica 1, cosine similarities are mostly greater than 0.8 with the sole exception of replica 5, where it is about 0.7.
More variability is observed for eigenvectors 8 and 9, which are either mixed (large diagonal and off-diagonal entries in the comparison matrix for replica 2), observed in a different order (replica 3), or not accurately reproduced at all (replica 5). 
Only for one case (replica 4), we observe essentially the same eigenvectors as in replica 1 with cosine similarities on the order of 0.9.

Increasing the length of simulation trajectories by a factor of two to 20~ns results in highly reproducible low-frequency vibrations.
While the order of eigenvectors describing translations and rotations still varies between distinct replicas, the lowest frequency vibrations are uniformly reproduced with correct ordering and correlations of 0.8 and higher (0.88 on average). 
Even eigenvectors with higher indices, which are expected to be increasingly influenced by noise as discussed earlier, are reproduced with high fidelity. 
A single exception are eigenvectors 11 and 12 in replica 3 whose order is swapped compared to replica 1.

Our results demonstrate that 20~ns simulations are sufficient to obtain reliable information on the lowest-frequency vibrations of HEWL. 
The latter can be explained by the rapid convergence of velocity time correlation functions, which typically decay to zero on timescales of 1-2~ps. 
Thus, simulation trajectories of 20~ns length contain between 10,000 and 20,000 independent samples.
The latter is not the case for methods reliant on correlated displacements, {\em i.e.}, coordinate fluctuations, which typically feature much longer correlation times and thus suffer from reduced statistics.

\section{Conclusion}
\label{s:con}

The analysis of molecular vibrations at low frequencies has long been limited by harmonic approximations, which are either used explicitly (harmonic normal mode analysis) or implicitly, {\em i.e.}, whenever a one-to-one mapping between DOFs and vibrational modes is assumed.
Harmonic approximations can provide a sound basis for the description of molecular vibrations at high frequencies, where quantum oscillators primarily populate their vibrational ground state and even classical oscillators are restricted to the immediate vicinity of a potential energy minimum.
However, at low frequencies, where the thermal energy significantly exceeds the harmonic oscillator excitation energy ($k_B T > h \nu$), harmonic approximations become increasingly invalid.
FRESEAN mode analysis \cite{sauer2023} provides a novel perspective on low-frequency vibrations in molecules, which does not rely on harmonic approximations.
Instead of mapping all degrees of freedom to a single set of orthogonal vibrational coordinates, dynamic fluctuations in a simulation trajectory are analyzed as a function of frequency.
As a result, orthogonal sets of collective vibrational coordinates are generated for each sampled frequency and sorted by their contributions (kinetic energy) to the vibrational spectrum at that frequency.

Thus, the simulated dynamics of the molecular system determines how many vibrational modes contribute to the spectrum at any given frequency.
Spectral lineshapes, damping, and the influence of the molecular environment are explicitly taken into account and can be analyzed via projections of the simulated dynamics onto isolated vibrational coordinates. 
Notably, the lowest frequency vibrations of large biomolecules such as proteins can provide hints on intrinsic collective degrees of freedom associated with conformational transitions, which can serve as effective CVs in enhanced sampling simulations.

However, the calculation of the velocity time cross correlation matrix, which is central to FRESEAN mode analysis, can be problematic for all degrees of freedom of a large system.
To sample all vibrational frequencies and avoid aliasing in the frequency domain, trajectory information needs to be stored (or processed on-the-fly, {\em e.g.}, via trajectory streaming~\cite{IMDClient}) at a high time resolution.
More importantly, the dimensions of the correlation matrix, whose elements consist of arrays of time correlations, can become prohibitive for large systems.

However, for many applications, such as the generation of CV candidates for enhanced sampling simulations, only collective vibrations at the lowest frequencies are of primary importance. 
Thus, we analyzed here the persistence of information on such low-frequency vibrations in distinct CG schemes applied {\em a posteriori} to all-atom simulation trajectories prior to analysis.
We find that CG schemes that map individual atoms of amino acids residues to beads describing their center-of-mass motion (one or two beads per residue), successfully capture essentially all information on collective frequency vibrations that give rise to the eigenvectors of the velocity correlation matrix at zero frequency.
Likewise, such CG representations serve as a natural low-pass frequency filters, which reduce the time resolution needed to prevent aliasing during Fourier analysis.

In addition, we analyzed the length of simulation trajectories needed to obtain highly reproducible results. 
The latter may be system-dependent but we expect our results to be applicable to proteins of similar size as HEWL.
Our optimal simulation protocol utilizes either a two-bead or one-bead (per residue) CG representation of the protein with a time resolution of 20~fs or 40~fs, respectively, and simulation trajectories of 20~ns length.

In Table~\ref{t:comp}, we show the relative computational costs (floating point operations = flops) and memory requirements associated with FRESEAN mode analysis of a 20~ns trajectory of HEWL (1960 atoms) for distinct representations.
For each representation, we use a time resolution of the trajectory that minimizes aliasing: 8~fs for all-atom, 20~fs for the two-bead representation (246 beads total), and 40~fs for the one-bead representation (129 beads total). 
Here, we assume that the elements of the time correlation matrix are computed from a single trajectory using the approach described in Eq.~\ref{e:conv}.

\begin{table} [ht!]
\centering
\begin{tabular}{ l | c | c | c }
 & all-atom & two-bead & one-bead \\ \hline \hline
 store trajectory (memory) & 100\% & 12.5\% & 6.6\% \\ 
 compute $\vc{C}_\tx{\tilde{v}}(\omega)$ (flops) & 100\% & 0.59\% & 0.08\% \\
 store $\vc{C}_\tx{\tilde{v}}(\omega)$ (memory) & 100\% & 0.63\% & 0.09\% \\
 diagonalize $\vc{C}_\tx{\tilde{v}}(0)$ (flops) & 100\% & 0.02\% & 0.003\%
\end{tabular}
\caption{Comparison of computational cost and memory requirements associated with distinct steps of FRESEAN mode analysis for 20~ns simulation of HEWL in distinct representations.
The number of particles are 1960 (all-atom), 246 (two-bead), and 129 (one-bead).
The time resolution of the trajectories for the distinct representations are 8~fs (all-atom), 20~fs (two-bead), and 40~fs (one-bead).
}
\label{t:comp}
\end{table}

Table~\ref{t:comp} clearly illustrates how dramatically computational costs and memory requirements decrease when the analysis is performed for CG representations of all-atom trajectories. 
This makes it straightforward to apply FRESEAN mode analysis to extract the lowest-frequency vibrations even for large proteins and protein complexes.

\raggedbottom

\begin{acknowledgements}
This work is supported by the National Science Foundation (CHE-2154834) and the National Institute of General Medical Sciences (R01GM148622). The authors acknowledge Research Computing at Arizona State University for providing high performance computing resources that have contributed to the research results reported within this work.
\end{acknowledgements}

%
%
\pagebreak
\begin{widetext}
\newpage
\begin{center}
\textbf{\large Supporting Information: High-Throughput Computation of Anharmonic Low-Frequency Protein Vibrations}
\end{center}
\setcounter{equation}{0}
\setcounter{figure}{0}
\setcounter{table}{0}
\setcounter{page}{1}
\makeatletter
\renewcommand{\theequation}{S\arabic{equation}}
\renewcommand{\thefigure}{S\arabic{figure}}
\renewcommand{\thetable}{S\arabic{table}}
\renewcommand{\thepage}{S\arabic{page}}

\subsection*{Eigenvector Comparison for One-Bead Representation}

\begin{figure}[ht!]
\centering
\includegraphics[width=1\textwidth]{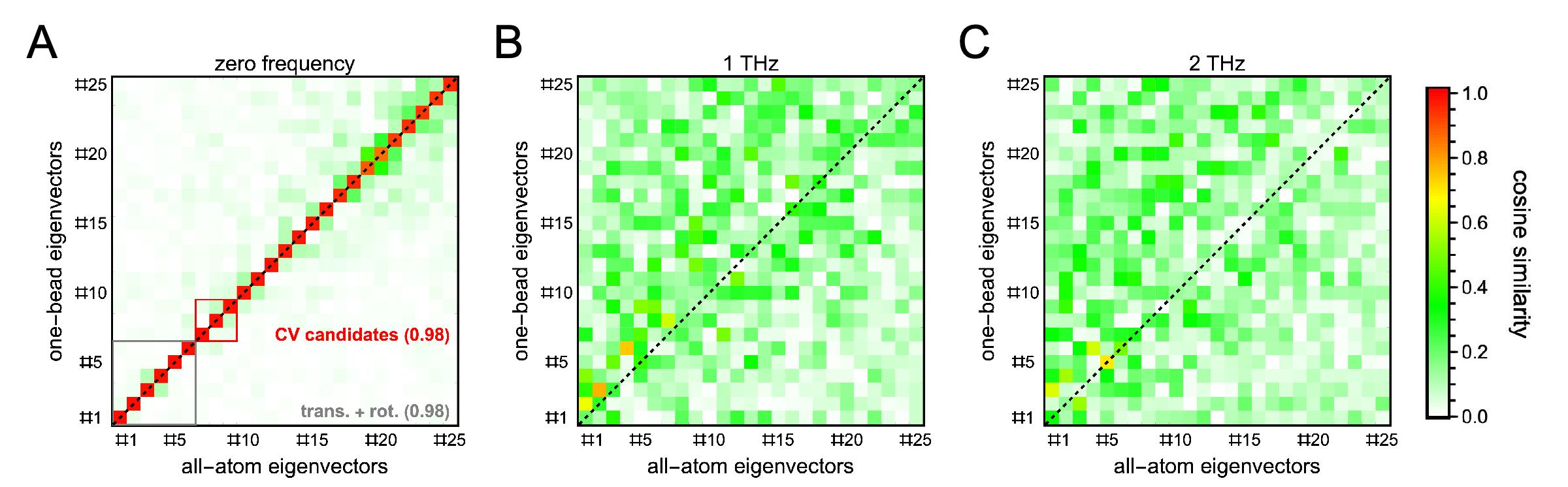}
\caption{Cosine similarity matrices between the first 25 eigenvectors of the velocity correlation matrix $\vc{C}_\tx{\tilde{v}}(\omega)$ at zero frequency (A), 1~THz (B), and 2~THz (C). 
Each matrix compares eigenvectors sampled from 20 ns simulations. 
A cosine similarity of 1 indicates parallel eigenvectors, while a value of 0 indicates orthogonal eigenvectors. 
At zero frequency, we highlight eigenvectors describing rigid-body translations and rotations (eigenvectors 1-6, gray box) and vibrations with large contributions to the zero frequency VDoS (eigenvectors 7, 8, and 9, red box), which we identified as potential CV candidates for enhanced conformational sampling in previous work~\cite{Sauer2024FastSample}. Average cosine similarities along the diagonal within each box are indicate as numerical values.}
\end{figure}

\newpage

\subsection*{Translational and Rotational Eigenvectors}
\begin{figure}[ht!]
\centering
\includegraphics[width=0.75\textwidth]{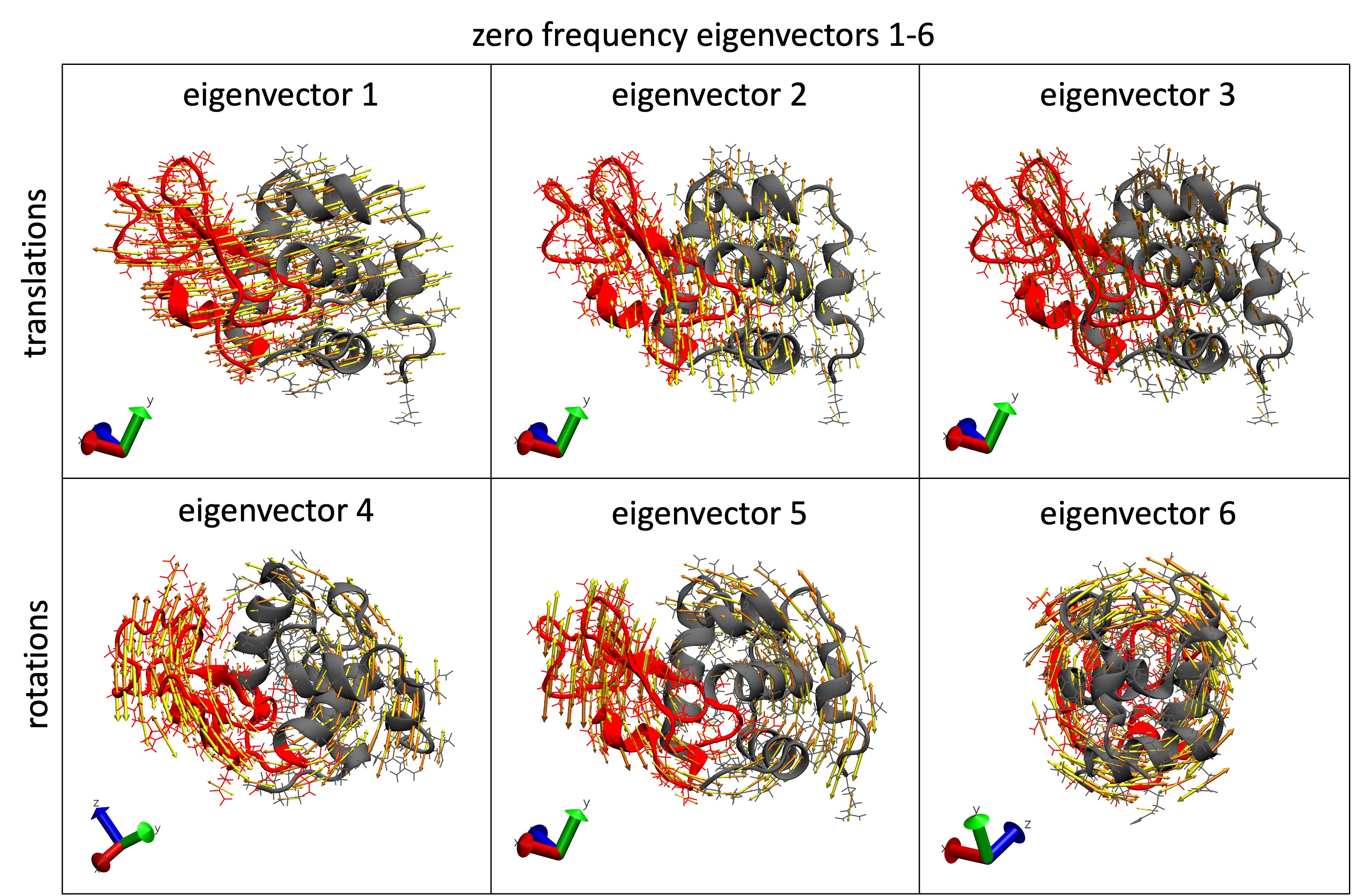}
\caption{Visualization of translational and rotational eigenvectors 1 - 6 of $\vc{C}_\tx{\tilde{v}}(\omega)$ at zero frequency obtained from all-atom trajectories. 
Displacement vectors are shown as bi-directional arrows that are superimposed on a reference structure. Orange (yellow) arrows describe the positive (negative) phase of an oscillation. The reference structure is shown in a cartoon representation to highlight secondary structure elements. The $\alpha$-domain is shown in gray and the $\beta$-domain is highlighted in red.}
\end{figure}

\newpage

\subsection*{Comparing Zero-Frequency Eigenvalues of Different Representations}
\begin{figure}[ht!]
\centering
\includegraphics[width=0.5\textwidth]{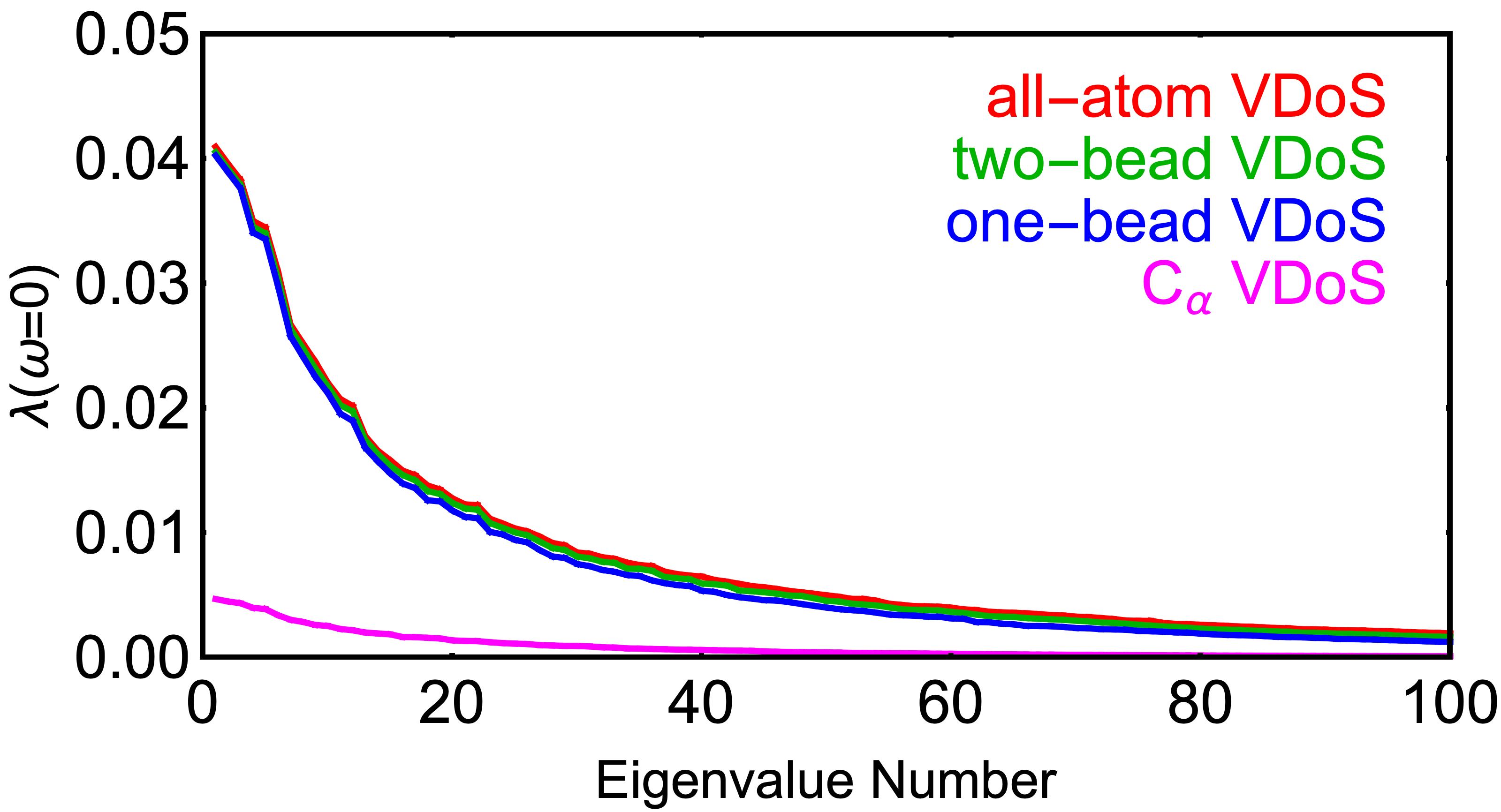}
\caption{First 100 eigenvalues at zero frequency obtained from the all-atom (red), two-bead (green), one-bead (blue), and \ca (magenta) representations.}
\end{figure}

\newpage

\subsection*{1D-VDoS Comparison with One Bead Model}
\begin{figure}[ht!]
\centering
\includegraphics[width=0.75\textwidth]{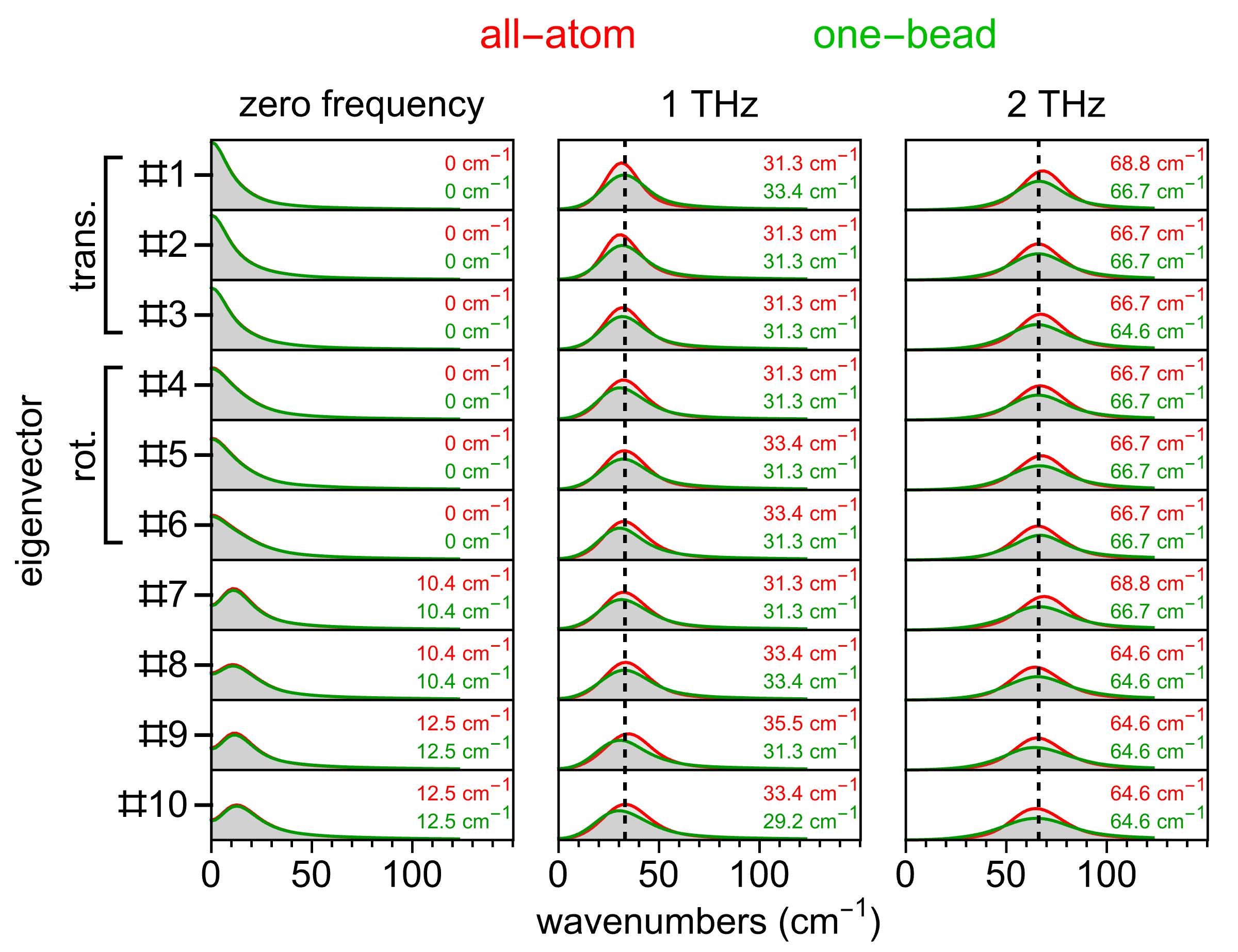}
\caption{1D-VDoS from fluctuations along the first ten eigenvectors of $\vc{C}_\tx{\tilde{v}}(\omega)$ at zero frequency, 1~THz, and 2 THz obtained for all-atom (red) and one-bead (blue) representations of protein trajectories. 
Colored labels indicate peak frequencies in each spectrum and vertical dashed lines indicate the frequency for which $\vc{C}_\tx{\tilde{v}}(\omega)$ was evaluated.}
\end{figure}

\newpage

\subsection*{Reducing Correlation Function Time Resolution for One-Bead Model}
\begin{figure}[ht!]
\centering
\includegraphics[width=0.5\textwidth]{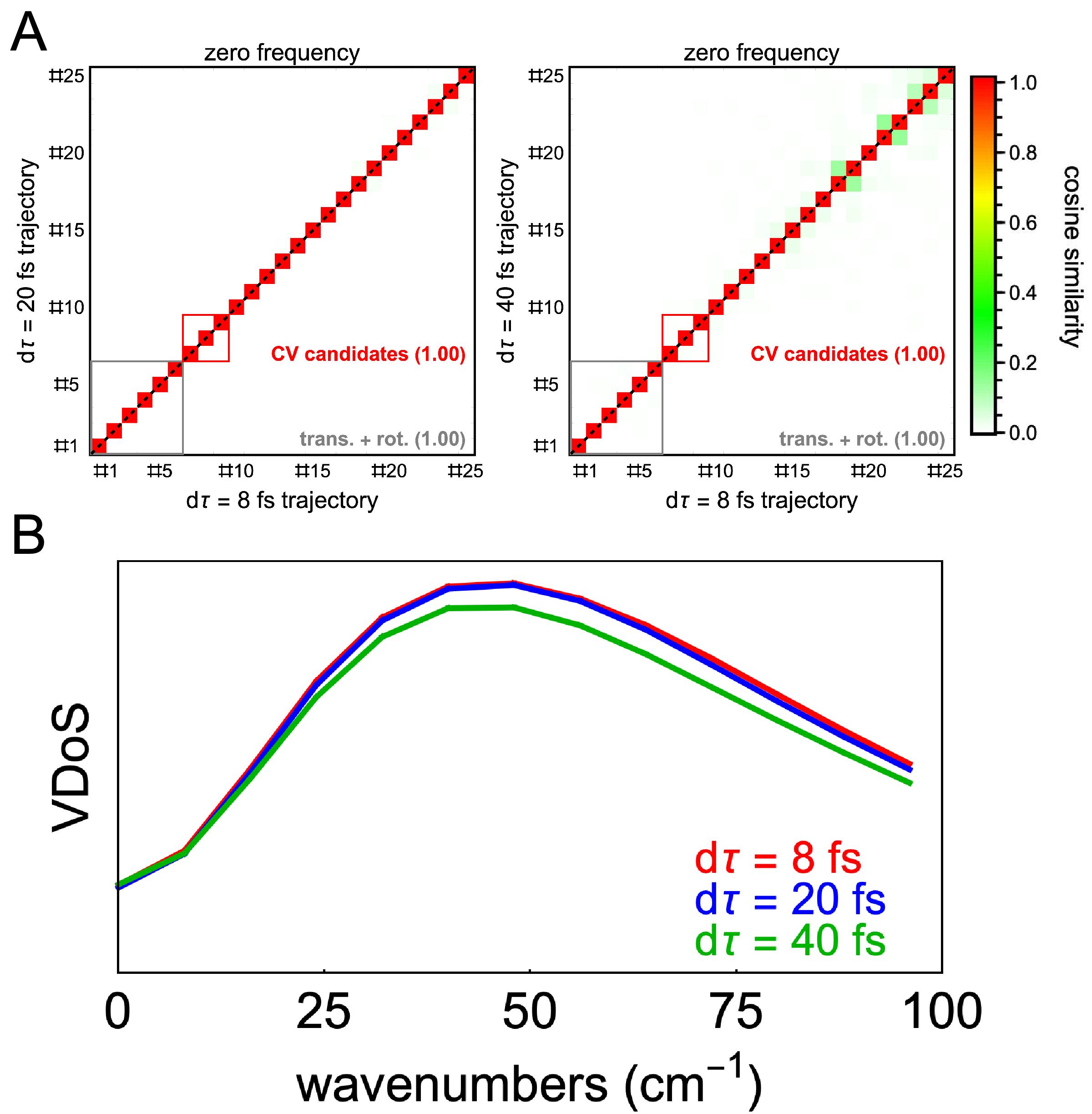}
\caption{(A) Cosine similarity matrices between the first 25 eigenvectors of the velocity correlation matrix $\vc{C}_\tx{\tilde{v}}(\omega)$ obtained with the one-bead representation at time resolutions of 8~fs and 20~fs (left) and 8~fs and 40~fs (right).
(B) VDoS obtained for the one-bead representation of trajectories with time resolutions of $d\tau$ = 8 (red), 20 (blue), and 40 fs (green).}
\end{figure}

\newpage

\subsection*{Evaluating Reproducibility of One-Bead modes at Different Simulation Times}
\begin{figure}[ht!]
\centering
\includegraphics[width=1\textwidth]{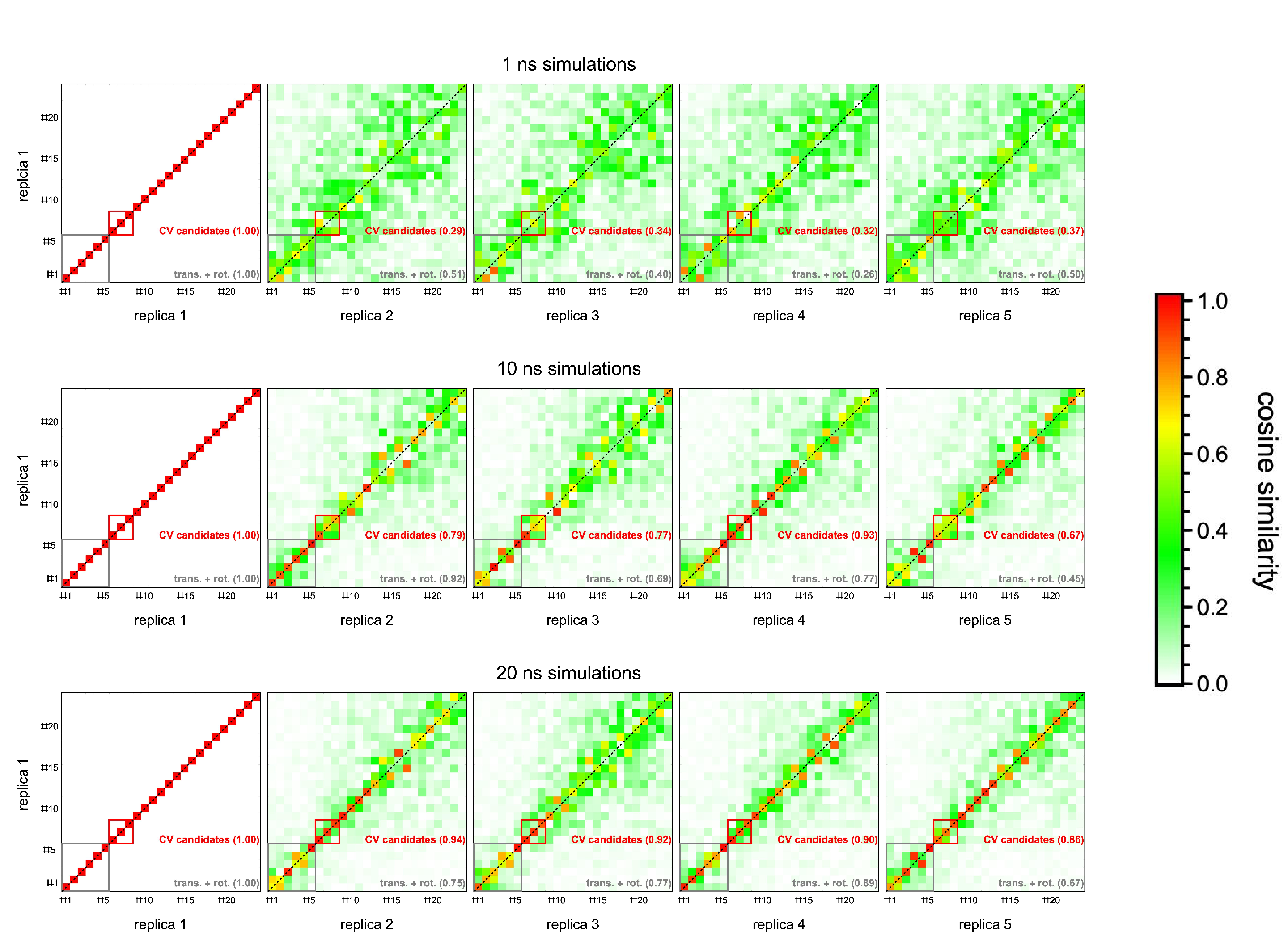}
\caption{Reproducibility of eigenvectors of $\vc{C}_\tx{\tilde{v}}(\omega)$ at zero frequency obtained from a one-bead representation of five replica all-atom simulations trajectories of 1~ns (A), 10~ns (B), and 20~ns (C) length.
All simulations are compared to replica 1 for clarity and shown are matrices of cosine similarities between pairs of the first 25 eigenvectors.
Eigenvectors describing rigid-body translations and rotations (1-6) and the lowest frequency vibrations (7-9) are highlighted in gray and red, respectively.}
\end{figure}

\end{widetext}

\newpage
%

\end{document}